\documentstyle[aps,prl,psfig,twocolumn]{revtex}
\begin{document}
\draft

\twocolumn[\hsize\textwidth\columnwidth\hsize\csname@twocolumnfalse\endcsname

\title{Low-frequency spin dynamics in the orthorombic phase of 
 $\bf La_{2}CuO_{4}$} 
\author{J. Chovan$^{1,2}$ and N. Papanicolaou$^{1,}$ \cite{NP_e-mail}}
\address{$^{1}$Department of Physics, University of Crete, and 
Research Center of Crete, Heraklion, Greece}
\address{$^{2}$Faculty of Science, P.J.\v{S}af\'{a}rik University,
         Park Angelinum 9, 04154 Ko\v{s}ice, Slovakia }

\maketitle
\date{\today}

\begin{abstract}
{\bf Abstract.}
An effective field theory is derived that describes the low-frequency spin
dynamics in the low-temperature orthorombic phase of $La_{2}CuO_{4}$.
Restricted to a single $CuO_{2}$ layer the effective theory is a simple
generalization of the relativistic nonlinear $\sigma$ model to include
all spin interactions allowed by symmetry. Incorporating a weak interlayer
interaction leads to two coupled nonlinear $\sigma$ models which provide
an efficient description of the complete bilayer dynamics.
Particular attention is paid to the weak-ferromagnetic and spin-flop
transitions induced by external magnetic fields. The main features of the
observed (covert) weak ferromagnetism are thus accounted for in a
straightforward manner but some of the finer theoretical predictions would
require further experimental investigation. The derived framework is also
suitable for the study of the structure and dynamics of magnetic domains in
undoped $La_{2}CuO_{4}$.
\end{abstract}


\vskip2pc
]

\tighten

\section{Introduction}

\label{sec:intro}

The magnetic properties of $La_{2}CuO_{4}$ have been extensively studied 
during the last decade \cite{1}.
In a crude approximation this system is described by a two-dimensional (2D)
isotropic Heisenberg antiferromagnet. However the orthorombic distortion of
the crystal that takes place below $530 K$ induces anisotropic spin couplings,
the most important of which is a Dzyaloshinskii-Moriya (DM) anisotropy 
\cite{2,3} that should lead to spin canting and thus weak ferromagnetism.
Yet a small antiferromagnetic interlayer coupling forces successive $CuO_{2}$
layers to cant in opposite direction and the induced weak moments
average to zero in the absence of external fields. Nevertheless the sustained
experimental as well as theoretical effort of Thio et al. \cite{4,5,6} 
established that $La_{2}CuO_{4}$ is indeed a covert weak ferromagnet.

The derived phenomenological picture was further probed by symmetry analysis
\cite{7,8} and by a fresh and challenging look at the DM anisotropy due to
Kaplan \cite{9} and Shekhtman, Entin-Wohlman, and Aharony \cite{10,11}.
The main outcome of the latter work is sometimes referred to as the KSEA
anisotropy and is the subject of current experimental investigation in the
related context of helimagnetism \cite{12}.

Our purpose is to systematize the above developments into a simple 
field theoretical framework that should facilitate further work on this
interesting subject. The relevance of effective field theories became apparent
through standard hydrodynamic approaches \cite{13,14} which eventually led to
a successful description of the isotropic Heisenberg antiferromagnet in terms
of a relativistic nonlinear $\sigma$ model \cite{15,16}.
A similar approach has been employed for the study of the dynamics of domain
walls and related topological magnetic solitons in conventional weak 
ferromagnets \cite{17}.  This issue has remained largely unexplored in the 
cuprates, 
apparently due to the hidden nature of weak ferromagnetism and a corresponding
lack of a complete field theoretical description.

 In Section 2 we repeat the symmetry analysis to obtain the most general spin
Hamiltonian involving nearest-neighbor (nn) interactions. Based on this 
Hamiltonian we proceed with the derivation of an effective field theory valid
at low frequencies in two steps. First, in Section 3, we derive a suitable 
extension of the 2D nonlinear $\sigma$ model which would be appropriate for
the description of a single $CuO_{2}$ layer provided that the interlayer 
coupling
could be neglected. Second, in Section 4, we incorporate a weak interlayer
interaction to obtain the final effective theory in the form of two coupled
nonlinear $\sigma$ models. The main phenomenological implications are also
worked out in Section 4, including a discussion of the observed 
weak-ferromagnetic \cite{4} and spin-flop \cite{5} transitions.
Our conclusions are summarized
in Section 5 by listing those issues that seem to deserve further attention.
In the Appendix we derive the most general next-nearest-neighbor (nnn)
in-plane interaction compatible with symmetry.

\section{symmetry constraints}

\label{sec:symm_con}
The crystal structure of $La_{2}CuO_{4}$ has been discussed on several
occasions and the accumulated information is freely used in this section.
In particular, we found the expositions of references \cite{18,19} lucid
and informative. The crystal undergoes a structural phase transition from
a tetragonal ($I4/mmm$) phase at high temperatures to an orthorombic
($Bmab$) phase below $530 K$. Throughout this paper we confine attention
to the low-temperature orthorombic ($LTO$) phase.

The relevant space group $Bmab$ is usually listed as $Cmca$ in standard
tables of crystallography using a slightly different choice of conventions.
Our conventions are illustrated in Figure 1 which depicts the orthorombic
unit cell stripped of all but the magnetic sites; i.e., the positions of the
spin $s=1/2$ $Cu^{2+}$ ions. Dashed lines in this figure join nn neighbors
within each $CuO_{2}$ plane and are the descendants of the original
tetragonal axes which are no longer orthogonal due to the orthorombic
distortion. In a first approximation, the magnetic ground state is such that
spins at sites denoted by $A$ and $\Delta$ point along the positive (negative)
$b$ axis, while spins at $B$ and $\Gamma$ point along the negative (positive)
$b$ axis \cite{20}. However, when anisotropies and a weak interlayer coupling
are taken into account, each spin suffers a slight canting which leads to
four inequivalent magnetic sites and a corresponding four-sublattice picture.
In the following, spin vectors are denoted by their standard symbol $\bf S$,
or by $\bf {A},\bf {B},\bf {\Gamma},\bf {\Delta}$ when a distinction among
the four sublattices becomes necessary.

\begin{figure}
\centerline{\hbox{\psfig{figure=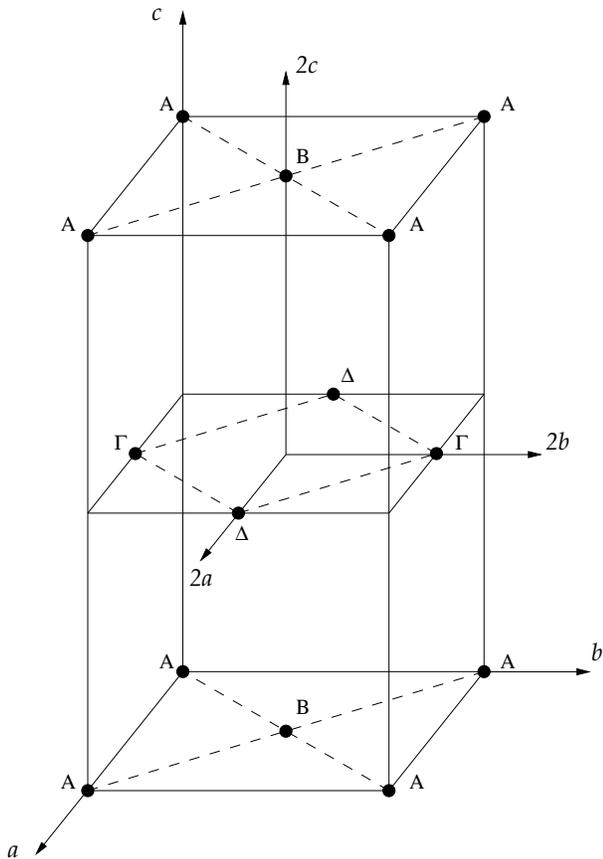,width=8cm}}}
\vskip0.5cm
\caption{The orthorombic unit cell of $La_{2}CuO_{4}$ stripped of all
but the $Cu$ atoms denoted by solid circles. Dashed lines join nn
neighbors within each $CuO_{2}$ plane and are the descendants 
of the original tetragonal axes.The four inequivalent magnetic sites 
are labeled by $A, B, \Gamma$ and $\Delta$.}
\end{figure}

In order to list the symmetry elements of the space group we begin with 
the most general primitive translation

\begin{equation}
 \label{eq:1}
 {\bf T}={\alpha}a{\bf e}_a + {\beta}b{\bf e}_b +
         {\gamma}c{\bf e}_c ,
\end{equation}
where {\bf e}$_a$, {\bf e}$_b$ and {\bf e}$_c$ are unit vectors along the
crystal axes, $a=5.35\AA$,  $b=5.40\AA~$ and $c=13.15\AA~$ are the
lattice constants, and ($\alpha, \beta, \gamma$) is a set of integers that
may also be used to label the relative position of a unit cell through 
the Cartesian coordinates $x={\alpha}a, y={\beta}b$ and $z={\gamma}c$.
We further consider the two fractional translations

\begin{equation}
 \label{eq:2}
 {\tau}=\frac{1}{2}(a{\bf e}_a + c{\bf e}_c) ,\hspace*{1cm}
 {\tau^{\prime}}=\frac{1}{2}(a{\bf e}_a + b{\bf e}_b) ,
\end{equation}
where $\tau$ is in itself a symmetry operation. The symmetry group of
the unit cell is then written symbolically as
 $G=G_{0}+{\tau}{\times}G_{0}$ where $G_{0}$ contains the eight elements

\begin{equation}
 \label{eq:3}
    E, I,\sigma_{a} , \sigma^{\prime}_{b} ,\sigma^{\prime}_{c} ,C_{2a} ,
    C^{\prime}_{2b} ,C^{\prime}_{2c} . 
\end{equation}
Here $E$ denotes identity, $I$ inversion about a Cu site, $ {\sigma}_a ,
{\sigma}_b ,{\sigma}_c$ reflections about the planes $x=a/2,$ $y=b/2,$ $z=c/2,$
and $C_{2a} ,C_{2b} ,C_{2c}$ $180^{\circ}$
rotations around the axes that emanate
from the center of the unit cell as shown in Figure 1. Primed elements in 
equation (3) must be  complemented by the special fractional translation
 $\tau^{\prime}$ which is $\em {not}$ in itself a symmetry operation. 
Finally we note that all elements of $G$ can be generated from the
fundamental set ($\tau, I, {\sigma}_a , C^{\prime}_{2c}$) by suitable
group multiplications. The above four elements have been used to derive
the spin Hamiltonian described in the remainder of this section without
presenting the detailed symmetry arguments.

 We first discuss a 2D restriction that would be appropriate for the 
description of a single $CuO_2$ layer if the interlayer coupling could be 
neglected. The special symmetry element $\tau$ of equation (2) maps $AB$ bonds
in the basal plane to ${\Gamma}{\Delta}$ bonds in the middle plane.
Therefore it is sufficient to consider for the moment only the basal plane.
 We assume that the Hamiltonian contains terms that are at most quadratic in
the spin operators, even though quartic terms are occasionally invoked to 
explain some features of multimagnon dynamics \cite{21}. Including all
nn interactions within the plane the Hamiltonian is written as the sum
of four terms :

\begin{equation}
 \label{eq:4}
  W = W_{E} + W_{DM} + W_{A} + W_{Z} ,
\end{equation}
which we now describe in detail. The first term

\begin{equation}
 \label{eq:5}
  W_{E} =  \sum_{<kl>} J_{kl}({\bf S}_{k}\cdot {\bf S}_{l})   
\end{equation}
contains the isotropic exchange interaction over nn in-plane bonds
denoted by $<kl>$. Symmetry requires that

\begin{equation}
 \label{eq:6} 
  J_{kl} = J , \hspace*{1cm}  \rm {for\;all\;in{-}plane \; nn\ bonds}.
\end{equation}
The second term

\begin{equation}
 \label{eq:7} 
  W_{DM} = \sum_{<kl>}{\bf D}_{kl}\cdot({\bf S}_{k}\times{\bf S}_{l})
\end{equation}
is the standard (antisymmetric) DM anisotropy \cite{2,3}. The vectors
${\bf D}_{kl}$ are severely restricted by symmetry and take only two distinct
values:

\begin{equation}
 \label{eq:8} 
  {\bf D}_{1} = D{\bf e}_{a} + D^{\prime}{\bf e}_{b} ,\hspace*{1cm} 
  \;{\bf D}_{2} = D{\bf e}_{a} - D^{\prime}{\bf e}_{b} ,
\end{equation}
which are distributed over the 2D lattice as shown in Figure 2 where
a sign alternation $\pm{\bf D}_{1}$ and $\pm{\bf D}_{2}$ on opposite
bonds is also displayed. This alternation together with the specific
form of DM vectors (8) were derived by Coffey, Bedell, and Trugman \cite{7}.
Actually the DM
vectors given in the above reference differ from those
of equation (8) by a $45^{\circ}$ rotation, apparently because they were
referred to the original tetragonal axes. Since the latter axes are not
exactly orthogonal in the $LTO$ phase, statement (8) should be
viewed as slightly more precise. The third term

\begin{equation}
 \label{eq:9} 
  W_{A} = \frac{1}{2}\sum_{<kl>}\sum_{i,j}K^{ij}_{kl}
          (S^{i}_{k}S^{j}_{l} + S^{j}_{k}S^{i}_{l})
\end{equation}
encompasses all ``symmetric'' exchange anisotropies over nn in-plane
bonds. Again the matrices $K_{kl}$ are restricted by symmetry to two
possible values:

\begin{eqnarray}
 \label{eq:10} 
     && 
 K_{1}=\left (\begin{array}{ccc}
                  K_{aa} & K_{ab} & 0 \\
                  K_{ab} & K_{bb} & 0 \\
                  0      & 0      & K_{cc} 
         \end{array} \right ) ,
             \nonumber
             \\
           & &
             \nonumber
             \\
           & &
             \nonumber
             \\  
      &&
 K_{2}=\left (\begin{array}{ccc}
                K_{aa} & \hspace*{-0.1cm} -K_{ab}  & 0 \\
               \hspace*{-0.1cm}-K_{ab}  & K_{bb} & 0 \\
                0      & 0      & K_{cc}
          \end{array} \right ) ,
\end{eqnarray}
which are distributed as shown in Figure 2. The above matrices may be
taken to be traceless ($K_{aa} + K_{bb} + K_{cc} = 0$) because the isotropic
component of the exchange interaction has already been accounted for by
equation (5).
Finally the fourth term

\begin{equation}
 \label{eq:11}
  W_{Z}= -\sum_{l}({\bf H}\cdot{\bf S}_{l})
\end{equation}
is simply the Zeeman interaction produced by an external field $\bf{H}$.

The symmetry analysis was further extended to include all nnn in-plane
interactions; namely, couplings along the diagonals of the $Cu$ plaquettes
which are parallel to the orthorombic $a$ and $b$ axes.
The resulting additions to the Hamiltonian are summarized in the Appendix.
However, although we have kept track of the nnn interactions throughout
our analysis, the corresponding results will not be included in the main
text because we wish to keep the exposition reasonably simple. But we will
occasionally mention potential contributions from the nnn interactions.

We next turn our attention to possible microscopic mechanisms that produce the
various anisotropies. A good starting point is the KSEA Hamiltonian
\cite{11}
\begin{figure}

\centerline{\hbox{\psfig{figure=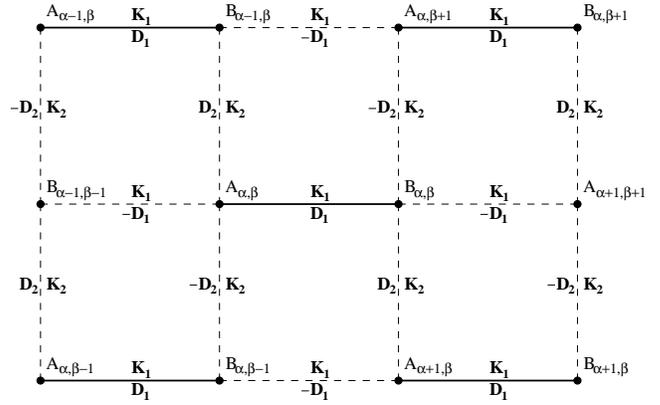,width=8.5cm}}}
\vskip1pc
\caption{ Illustration of the distribution of the DM vectors $\pm {\bf D}_{1}$
and $\pm {\bf D}_{2}$, and of the symmetric exchange matrices $K_{1}$
and $K_{2}$, on a portion of the basal plane. The spins ${\bf A}$ and
${\bf B}$ in a given dimer are labeled by a pair of indices 
$({\alpha}, {\beta})$ that advance along the orthorombic axes $a$ and $b$
not shown in the figure.}
\end{figure} 
\begin{eqnarray}
 \label{eq:12}
       W &=&\sum_{<kl>}[\;(J_{kl} - \frac{|{\bf D}_{kl}|^2}{4J_{kl}})
                   ({\bf S}_{k}\cdot{\bf S}_{l}) + 
                    {\bf D}_{kl}\cdot({\bf S}_{k}\times{\bf S}_{l})
     \nonumber
      \\
        & &+ \frac{1}{2J_{kl}}({\bf S}_{k}\cdot{\bf D}_{kl})
                        ({\bf D}_{kl}\cdot{\bf S}_{l})\;]  ,
\end{eqnarray}
obtained by carrying the perturbative treatment of spin-orbit interaction
to second order. The second-order terms in equation (12) are usually
neglected by comparison to the linear DM anisotropy. However their effect can
be far more subtle noting that the total contribution from each specific mode
can be brought to a completely isotropic form by rotating the spin
operators ${\bf S}_{k}$ and ${\bf S}_{l}$ around the ${\bf D}_{kl}$ axis
with angles $-\theta_{kl}$ and $\theta_{kl}$ such that 
$\tan\theta_{kl}=|{\bf D}_{kl}|^{2}/2J_{kl}$. One would naively conclude
that Hamiltonian (12) does not predict spin ordering of any kind;
in particular, weak ferromagnetism. Nevertheless spin ordering may
occur as a result of frustration when the collective effect from all bonds
is taken into account. This delicate line of reasoning \cite{11} led to
some interesting predictions such as the existence of weak ferromagnetism
in the $LTO$ phase of pure $La_{2}CuO_{4}$ but its absence in the 
low-temperature tetragonal ($LTT$, space group $P4_{2}/ncm$) phase that
may be realized in, say, $Ba$-doped $La_{2-x}Ba_{x}CuO_{4}$.

It should be expected that the exchange anisotropy abstracted from
equation (12) is consistent with the symmetry statement (10). Indeed, taking
into account that $J_{kl}=J$ for all in-plane bonds and the specific form
of the DM vectors from equation (8), the (traceless) KSEA anisotropy is found to be
a special case of (10) with

\begin{eqnarray}
 \label{eq:13}
       &&      
  K_{aa} = \frac{2D^{2}-D^{\prime2}}{6J} , \hspace{1cm}
  K_{bb} = \frac{2D^{\prime2}-D^{2}}{6J} , 
       \nonumber
       \\
       &&
       \nonumber
       \\
       &&
  K_{cc} = - \frac{D^{2}+D^{\prime2}}{6J} , \hspace{1cm}
  K_{ab} = \frac{DD^{\prime}}{2J} ,
\end{eqnarray}
Of course, spin-orbit interaction is not the only source of anisotropy and,
in fact, Coulomb exchange produces an Ising-like term \cite{11,22}
described by

\begin{equation}
 \label{eq:14}
  K_{aa} = \frac {1}{3}K = K_{bb},\hspace*{1cm} K_{cc} = -\frac {2}{3}K,
\end{equation}
which should be added to the corresponding elements of equation (13).
Therefore we proceed with caution using the most general anisotropy
given by equation (10) and return to further discussion of the special
cases (13) and (14) after the dust settles, in Section 3.

The remainder of this section is devoted to a brief explanation of our
main strategy. At this point we invoke the classical approximation, 
the validity of which will not be questioned until Section 4.
The 2D dynamics will thus be described by the Landau-Lifshitz 
equation \cite{23} in the form given by Gilbert \cite{24}:
        
\begin{eqnarray}
\label{eq:15}
 \frac {\partial{\bf A}}{\partial t} +
 \gamma ({\bf A}\times \frac{\partial {\bf A}}{\partial t}) &=& 
            {\bf A}\times {\bf F}_{A},\hspace*{0.5cm} {\bf F}_{A} = 
              -\frac{\partial W}{\partial {\bf A}} ,
        \nonumber
        \\  
        & &
        \nonumber
        \\                                       
 \frac{\partial{\bf B}}{\partial t} +
 \gamma ({\bf B}\times \frac{\partial {\bf B}}{\partial t}) &=& 
            {\bf B}\times {\bf F}_{B},\hspace{0.5cm} {\bf F}_{B} = 
              -\frac{\partial W}{\partial {\bf B}} ,    
\end{eqnarray} 
where $\gamma$ is a dissipation constant. The spin variables ${\bf A}$ and
${\bf B}$ are treated as classical vectors of length $s$, and the lattice
indices ($\alpha,\beta$) displayed in Figure 2 are suppressed in equation
(15) for notational convenience.

The suppression of indices is more than symbolic if one wishes to study
only the homogeneous spin dynamics in the presence of a spatially uniform
field ${\bf H}$. Then spins assume only two distinct values ${\bf A}$ and
${\bf B}$, one for each sublattice, and the effective fields ${\bf F}_{A}$
and ${\bf F}_{B}$ in equation (15) may be derived from the much simpler
Hamiltonian
\begin{eqnarray}
  \label{eq:16}
   W_{C} &=& 4[ J({\bf A}\cdot{\bf B}) + D(A_{b}B_{c} - A_{c}B_{b}) 
          \nonumber
          \\
         & & + K_{aa}A_{a}B_{a} + K_{bb}A_{b}B_{b} + K_{cc}A_{c}B_{c}]
          \nonumber
          \\
         & & - {\bf H}\cdot({\bf A}+{\bf B}) ,
\end{eqnarray}
where ($ A_{a},A_{b},A_{c}$) and ($B_{a}, B_{b}, B_{c}$) 
are the Cartesian components
of the spin vectors ${\bf A}$ and ${\bf B}$ along the orthorombic axes.
A notable fact is that neither $D^{\prime}$ nor $K_{ab}$ enter equation (16).
If we further restrict the diagonal anisotropies to the Ising form (14), 
equation (16) yields the unit-cell Hamiltonian employed by Thio et al.
to account for a wide range of experiments \cite{4,5,6}. The sign alternation
of the DM vectors on opposite bonds \cite{7} is crucial for the description
in terms of a unit-cell Hamiltonian and was thus implicitly assumed in
reference \cite{4}. Lack of sign alternation would lead to spiral magnetic
order or helimagnetism \cite{12}.

The information accumulated so far is employed in Section 3 to study the 2D
dynamics of a single layer. One should recall that the 
dynamics of the middle plane is completely isomorphic and may be obtained
by the simple substitution
$(\bf {A},\bf {B})\rightarrow (\bf {\Gamma},\bf {\Delta})$.
The complete 3D dynamics including interlayer interactions will be studied
in Section 4.

\section{Dynamics of a single layer}

\label{sec:dyn_sing_lay}

The unit-cell Hamiltonian (16) may be analyzed through the classical
Landau-Lifshitz equations (15) to furnish explicit predictions for the
characteristic magnon frequencies and the corresponding dynamic
susceptibilities, following the early treatment of orthoferrites by Herrmann
\cite{25,26}. Considerable simplifications are invoked along the way
by appealing to the phenomenological fact that anisotropies and the
applied field are much smaller than the isotropic exchange constant:
\begin{equation}
 \label{eq:17}
   D, K, H \ll J.
\end{equation}
Simply stated our task is to derive an effective low-frequency dynamics
in which the strong inequalities (17) are taken into account from the outset.
And, at little extra cost, we aim to go beyond the homogeneous dynamics
to include spatial variations within a complete continuum field theory
in the form of a nonlinear $\sigma$ model.

One must thus deal with the complete discrete Landau-Lifshitz equations (15)
where $\bf {A}=\bf {A}_{\alpha,\beta}$ and $\bf {B}=\bf {B}_{\alpha,\beta}$ 
are the spin
vectors in a generic dimer labeled by two integers ($\alpha,\beta$) that
advance along the orthorombic $a$ and $b$ axes, as shown in Figure 2.
The corresponding effective fields ${\bf F}_{A}$ and ${\bf F}_{B}$ are
then constructed by closely following the prescription for the complete
microscopic Hamiltonian (4) given in Section 2.
The resulting algebraic expressions are somewhat lengthy but the actual
derivation is a straightforward adaptation of a direct method employed
earlier for the study  of an 1D model weak ferromagnet \cite{27} and a 2D
antiferromagnet \cite{28}. We thus suppress the algebraic details
and state the final result which can be explained in a simple manner.

Rationalized space-time variables are defined from 

\begin{equation}
 \label{eq:18}
     \eta = \alpha\delta ,\hspace*{1cm}
     \xi = \beta\delta ,\hspace*{1cm}
     \tau = 2s\delta Jt,
\end{equation}
where ${\delta}$ is a dimensionless scale whose significance will be made 
precise as the argument progresses. The actual distances along the $a$
and $b$ axes are then given  by $x= \eta a/\delta$ and $y= \xi b/\delta$
while frequency is measured in units of $2s\delta J$.
 We also define rescaled parameters grouped into two categories.
The DM anisotropy and the applied field are scaled linearly with $\delta$ :

\begin{equation}
 \label{eq:19}
     d = \frac{2D}{\delta J} ,\hspace*{1cm}
    {\bf h} = \frac{{\bf H}} {2s\delta J} ,     
\end{equation}
whereas diagonal anisotropies are scaled quadratically:

\begin{equation}
 \label{eq:20}
     \kappa_{a} = \frac{8K_{aa}}{\delta^{2} J} ,\hspace*{1cm}
     \kappa_{b} = \frac{8K_{bb}}{\delta^{2} J} ,\hspace*{1cm}   
     \kappa_{c} = \frac{8K_{cc}}{\delta^{2} J} .
\end{equation}
Note that the parameters $D^{\prime}$ and $K_{ab}$ do not appear in the
above list because they eventually drop out of the effective
low-frequency dynamics, for essentially the same reason they do not
appear in the unit-cell Hamiltonian (16).

Concerning the field variables a transparent formulation is obtained
in terms of the ``magnetization'' $\bf {m}$ and the ``staggered
magnetization'' $\bf {n}$ which are defined by

\begin{equation}
 \label{eq:21}
  {\bf m}= \frac{1}{2s}({\bf A} + {\bf B}) ,\hspace*{1cm}
  {\bf n}= \frac{1}{2s}({\bf A} - {\bf B}) ,
\end{equation}
and satisfy the classical constraints ${\bf m}\cdot{\bf n}=0$ and
${\bf m}^{2} + {\bf n}^{2} = 1$. The basis for the derivation of an
 effective field theory is that the strong inequalities (17)
imply $|{\bf m}|\ll|{\bf n}|$. Indeed, a consistent low-frequency
reduction of the Landau-Lifshitz equation is obtained by treating
${\bf m}$ as a quantity of order ${\delta}$, whereas the staggered
 moment ${\bf n}$ and the rescaled variables (19)-(21) are assumed
to be of order unity. Then, to leading order, the classical constraints
reduce to

\begin{equation} 
 \label{eq:22}
  {\bf m}\cdot{\bf n} = 0 ,\hspace*{1cm}
  {\bf n}^{2} = 1 ,
\end{equation}
and  ${\bf m}$ may be expressed in terms of {\bf n} through
the algebraic relation

\begin{equation}
 \label{eq:23}  
  {\bf m} = \frac {\delta}{4}[{\bf n}\times
            (\dot{\bf n} + {\bf d} - {\bf n}\times{\bf h})
            - ({\bf n}_{\eta} + {\bf n}_{\xi})] ,
\end{equation}
where ${\bf d}=d{\bf e}_{a}$, the dot denotes differentiation with respect
to $\tau$, and subscripts differentiation with respect to the spatial
coordinates $\eta$ and $\xi$. The various pieces of equation (23) have
appeared in earlier studies of conventional weak ferromagnets \cite{17,27}
 and antiferromagnets \cite{28} and will not be discussed here
in detail.

The main point is that the dynamical equations may now be stated
entirely in terms of the staggered moment ${\bf n}$ which satisfies
the nonlinear $\sigma$ model

\begin{equation}
 \label{eq:24}  
  {\bf n}\times({\bf f} + \lambda \dot{\bf n}) = 0 ,\hspace*{1cm}
  {\bf n}^{2} = 1 ,
\end{equation}
where $\lambda = 4s\gamma /\delta$ is a rescaled dissipation
constant, and the effective field ${\bf f}$ may be derived from 
an action principle :
 
\begin{equation}
 \label{eq:25}
 {\bf f} = - \frac {\delta {\cal A}}{\delta {\bf n}},\hspace*{1cm}
             {\cal A} = \int {L\,d{\eta}d{\xi}d{\tau}} ,
\end{equation}
where ${\cal A}$ is the action and $L$ the corresponding Lagrangian
density

\begin{equation}
 \label{eq:26}
   L = L_{0} -  V.
\end{equation}
Here $L_{0}$ is the ``free'' Lagrangian

\begin{equation}
 \label{eq:27}
   L_{0} = \frac{1}{2}(\dot{\bf n}^{2}
           - {\bf n}^{2}_{\eta} -  {{\bf n}}^{2}_{\xi})
           + {\bf h}{\cdot}({\bf n}\times\dot{\bf n})
\end{equation}                                               
and $V$ is the ``potential''

\begin{equation}
 \label{eq:28}
                V = ({\bf h}\times {\bf d})\cdot{\bf n} + 
                \frac{1}{2}({\bf n}\cdot{\bf h})^{2} + 
                \frac{1}{2}(a^{2}_{1}n^{2}_{a} + 
                            a^{2}_{2}n^{2}_{c}) ,
\end{equation}
where ($n_{a}, n_{b}, n_{c}$) are the Cartesian components of ${\bf n}$
and

\begin{equation}
 \label{eq:29}
  a^{2}_{1} = d^{2} + \kappa_{b} - \kappa_{a} ,\hspace*{1cm}
  a^{2}_{2} = \kappa_{b} - \kappa_{c} ,
\end{equation}
are the final effective anisotropy constants.

If the applied field were absent (${\bf h}= 0$) the derived field theory
would be relativistically invariant. The ``velocity of light''
is equal to the magnon velocity in the corresponding isotropic
antiferromagnet and is scaled out of equation (27) thanks to the use
of rationalized units. Recalling that  $x= \eta a/\delta$, $y= \xi b/\delta$,
and  $\tau=2s\delta J$, the actual magnon velocities along the $a$ and $b$
directions are $V_{a} = 2saJ$ and $V_{b} = 2sbJ$. The
predicted slight anisotropy  $V_{b}/V_{a} = 1.01$
 cannot be discerned in current experiments which yield an average
spinwave velocity 
$V_{sw} \approx V_{a} \approx V_{b} \approx 850 meV\AA$
for pure $La_{2}CuO_{4}$ samples \cite{1}. Using an average lattice
 constant $a \approx b = 5.375 \AA$ one may extract a classical value for 
the exchange constant $J=158 meV$ which differs from the usually accepted
$J=135 meV$ by the calculated quantum-renormalization factor 1.18.

We are now in a position to make contact with the various special limits
of the exchange anisotropy discussed in Section 2.
The pure Ising anisotropy (14) leads to $a_{1} = d$ and 
$a^{2}_{2} = 8K/{\delta}^{2}J$ which correspond to the minimal model
of Thio et al. \cite{4}. In principle, the KSEA anisotropy (13)
could lead to $a_{1}\neq d$. However an estimate of
 the DM parameters
$D$ and $D^{\prime}$ within a tight-binding model \cite{11} yields
the near equality $|D|\approx |D^{\prime}|$ which simplifies
the diagonal elements of 
equation (13) to an Ising-like form, and thus  $a_{1} \approx d$,
whereas the off-diagonal element $K_{ab}$ does not appear
in the effective theory. Another mechanism for $a_{1}\neq d$
is provided by the diagonal nnn exchange anisotropies (see the Appendix)
but we do not have a way to theoretically estimate those parameters.
In short, departures from the strict equality $a_{1}= d$
should be expected but the near equality $a_{1} \approx d$ is
likely a good assumption. In any case, our theoretical calculations
will be carried out in terms of the three parameters 
($a_{1}, a_{2}, d$) and further discussion of this issue is
postponed to Section 4.3.

Next we discuss the field-dependent terms in equations (27) and (28).
First we note the well-known fact that the last term in the free
Lagrangian (27) breaks Lorentz invariance \cite{17}.
Incidentally, a mild breakdown of Lorentz invariance is also induced by
the off-diagonal nnn exchange anisotropy discussed in the Appendix.
The potential (28) contains two distinct contributions from the 
external field. The term $({\bf n}\cdot {\bf h})^{2}$ is an easy-plane
anisotropy due to the tendency of the two spins in a given dimer
to antialign in a direction nearly perpendicular to the applied field.
More subtle is the term $({\bf h}\times {\bf d})\cdot {\bf n}$ which
couples the external field to the antisymmetric DM anisotropy.
The existence of such a term was anticipated by Andreev and Marchenko
\cite{29} in their phenomenological treatment of conventional
weak ferromagnets based on symmetry. Although this term is often
dismissed in the literature \cite{17}, it was shown to be important
for a proper understanding of domain-wall dynamics \cite{27}.
In fact, the present work will provide ample evidence for the crucial
importance of such a term in every aspect of weak ferromagnetism.

Whereas the complete physical picture cannot be established until
we incorporate the interlayer coupling, in Section 4, the remainder
of this section is devoted to the derivation of some basic
consequences of the single-layer dynamics. Applications carried out
in this paper pertain to homogeneous spin dynamics. One may then neglect
spatial gradients to write

\begin{eqnarray}
\label{eq:30}
      {\bf m} &=& \frac{\delta}{4}
                  [{\bf n}\times (\dot{\bf n} + {\bf d}
                   - {\bf n}\times {\bf h})] ,
	\nonumber
	\\
       L_{0}  &=& \frac{1}{2}{\dot{\bf n}}^{2}  
                        + {\bf h}\cdot({\bf n}\times{\dot{\bf n}}) ,
	\nonumber
	\\
            V &=& ({\bf h}\times {\bf d})\cdot{\bf n} + 
                  \frac{1}{2}({\bf n}\cdot{\bf h})^{2} + 
                  \frac{1}{2}(a^{2}_{1}n^{2}_{a} + 
                            a^{2}_{2}n^{2}_{c}) ,
\end{eqnarray}
which describe the low-frequency dynamics associated with the unit-cell
Hamiltonian (16).

For our current demonstration we assume that the field points along the
$c$ axis ($h_{a} = 0 = h_{b}, h_{c} = h$) and thus the potential
reduces to

\begin{equation}
 \label{eq:31}
       V=hd{n}_{b}+
                \frac{1}{2}[a^{2}_{1}n^{2}_{a} +
                (a^{2}_{2} + h^{2})n^{2}_{c}],
\end{equation}
whose local minima are given by ${\bf n} = (0, \mp 1, 0)$ and the
corresponding magnetization is computed from the first equation
in (30) applied for static ${\bf n}$; i.e , $\dot{\bf n} = 0$. Hence the two
ground-state configurations are described by

\begin{equation}
 \label{eq:32}
   {\bf n} = \mp{{\bf e}_{b}} ,\hspace*{1cm}
   {\bf m} = \frac{\delta}{4}(h \pm d){\bf e}_{c}
\end{equation}
and are degenerate at zero field. For $h > 0$ the upper sign yields
the absolute ground state, whereas the lower sign corresponds to
a metastable local minimum with higher energy. Equation (32) makes
it explicit that a weak ferromagnetic moment develops along the $c$
axis even in the absence of an applied field.

The computation of the spectrum of normal frequencies is now
straightforward. In terms of the standard angular variables 

\begin{equation}
 \label{eq:33}
   {n}_{a}+i{n}_{b}=\sin \Theta\, e^{i \Phi} ,\hspace*{1cm}
   {{n}_{c}}=\cos \Theta ,
\end{equation}
the free Lagrangian is written as
 
\begin{equation}
 \label{eq:34}
          L_{0} = \frac{1}{2}
                 (\dot \Theta^{2} + \sin ^{2} \Theta\; \dot{\Phi}^{2})
                 + h\sin ^{2}{\Theta}\;\dot \Phi  ,
\end{equation}
and a corresponding angular parametrization of the potential is obtained
simply by inserting equation (33) in (31).

 The ground-state 
configurations are then given by $\Theta = \frac{\pi}{2},$
$\Phi = \mp \frac{\pi}{2}$, while small fluctuations are accounted
for by making the replacements $\Theta = \frac{\pi}{2} - \theta,$
$\Phi = \mp(\frac{\pi}{2} - \phi)$ and keeping terms that are at most
quadratic in ${\theta}$ and ${\phi}$. Also omitting a trivial additive
constant and total derivatives one finds that

\begin{equation}
 \label{eq:35}
          L = L_{0} - V \approx 
            \frac{1}{2}(\dot \phi^{2} + \dot \theta^{2}) - 
            \frac{1}{2}( \omega^{2}_{1^{\pm}} \phi^{2} + 
            \omega^{2}_{2^{\pm}} \theta^{2}),
\end{equation}
where

\begin{equation}
 \label{eq:36}
   \omega^{2}_{1^{\pm}} = a^{2}_{1} \pm hd ,\hspace*{1cm}
   \omega^{2}_{2^{\pm}} = a^{2}_{2} \pm hd + h^{2}
\end{equation}
are the (squared) characteristic magnon frequencies for in-plane and
out-of-plane fluctuations, respectively.

The actual experimental values of the calculated ``magnon gaps'' will
be discussed in Section 4. But first we must fill a gap in our
theoretical arguments concerning the choice of the scale parameter
${\delta}$ in equations (19)-(21). Although ${\delta}$ plays an
important
role in ascertaining the relative significance of the various terms
that arise during the low-frequency reduction of the Landau-Lifshitz
equations, consistency requires that all physical predictions be
independent of ${\delta}$. The magnon gaps (36) provide a good 
illustration of this point by recalling that the unit of frequency
is $2s{\delta}J$. When the right-hand sides of equations (36)
are multiplied by $(2s{\delta}J)^{2}$ the resulting expressions are
indeed independent of ${\delta}$ and contain only the original
 microscopic
parameters in suitable combinations. But one may exploit ${\delta}$
 to choose more convenient rationalized units as discussed in
Section 4.3.

This section is completed by quoting an explicit result for the
corresponding dynamic susceptibilities. The method of calculation
proceeds roughly as follows. Suppose that the system is found in 
its ground state in the presence of a constant field $h$ along
the $c$ axis. A weak AC field of arbitrary direction and frequency
${\omega}$ is then turned on and the system eventually relaxes into a
steady state oscillating with the same frequency thanks to the
dissipative term present in equation (24). Once the steady state
for the staggered moment ${\bf n}$ is determined from a perturbative
solution of the nonlinear ${\sigma}$ model, in the weak-AC-field limit,
the magnetization ${\bf m}$ is calculated from the first equation
in (30) to yield the sought after susceptibilities:
\begin{eqnarray}
 \label{eq:37}
    &&
    {\chi}_{aa} =  \frac{\delta}{4}
                    \left[
                    1+ \frac{{\omega}^{2}}
                    {{\omega}^{2}_{2} - {\omega}^{2} - i{\lambda}{\omega}}
                    \right] ,      
    \nonumber
    \\
    & &
    \nonumber
    \\
    &&
    {\chi}_{bb} =  \frac{\delta}{4}
                    \frac{(d + h)^{2}}
                   {{\omega}^{2}_{2} - {\omega}^{2} - i{\lambda}{\omega}} ,
    \nonumber
    \\
    & &
    \nonumber
    \\
    &&
    {\chi}_{cc} =  \frac{\delta}{4}
                    \left[
                    1+\frac{{\omega}^{2}}
                    {{\omega}^{2}_{1} - {\omega}^{2} - i{\lambda}{\omega}}
                    \right] ,       
    \nonumber
    \\ 
    & &
    \nonumber
    \\
    &&
    {\chi}_{ab} =  - {\chi}_{ba}	
                   = \frac{i\delta}{4}
                   \frac{(d + h){\omega}}
                   {{\omega}^{2}_{2} - {\omega}^{2} - i{\lambda}{\omega}} ,
    \nonumber
    \\
    & &
    \nonumber
    \\
    &&
    {\chi}_{ac} =  {\chi}_{ca} = {\chi}_{bc} =  {\chi}_{cb} = 0 ,
\end{eqnarray}
where $\omega_{1}$ and $\omega_{2}$ are the magnon gaps (36) restricted
to the upper (+) sign; i.e., the gaps above the absolute ground state
when $h>0$. The susceptibilities for the metastable ground state can be
obtained by the universal replacement $h \rightarrow -h$. Since the two
types of ground state mix in the presence of the interlayer coupling,
the results of equations (36) and (37) should be interpreted with
caution. 

\section{interlayer coupling}
\label{sec:int_coup}

Although interlayer interactions are expected to be weak, they are
important for a proper understanding of spin dynamics in $La_{2}CuO_{4}$.
We have thus extended the symmetry analysis of Section 2 to include
nn interlayer couplings on bonds that are parallel to either the
$ac$ or the $bc$ plane. For each $Cu$ atom there exist eight
out-of-plane neighbors, four in the plane above and another four
in the plane below. Symmetry requires that the isotropic exchange interaction
is described by an exchange constant $J_{1}$ for bonds that are parallel
to the $ac$ plane, and a second exchange constant $J_{2}$ for bonds
parallel to the $bc$ plane. These two constants would be equal in
the tetragonal ($I4/mmm$) phase but are different in the $LTO$ phase
due to the orthorombic distortion.

The exchange constants are expected to individually satisfy the
strong inequalities

\begin{equation}
 \label{eq:38}
  J_{1}, J_{2} \ll J ,
\end{equation}
in view of the fact that the length of out-of-plane nn bonds is significantly
larger than the length of in-plane nn bonds. It is thus reasonable to
assume that out-of-plane anisotropies (symmetric or antisymmetric) can
be safely ignored because they are expected to be even weaker. In fact,
we have worked out the form of all such anisotropies compatible with
symmetry to convince ourselves that they do not bring in potentially
new elements.

Therefore the three-dimensional (3D) unit-cell Hamiltonian can be
written as

\begin{equation}
 \label{eq:39}  
  W^{3D}_{C} = W_{C}({\bf A}, {\bf B}) + 
               W_{C}({\bf \Gamma}, {\bf \Delta}) +
               W_{int}({\bf A},{\bf B},{\bf \Gamma},{\bf \Delta}) ,
\end{equation}
where the first term is the 2D unit-cell Hamiltonian (16), the second
term is obtained by the simple substitution 
$(\bf {A},\bf {B})\rightarrow (\bf {\Gamma},\bf {\Delta})$,
and 
$W_{int}$ contains the isotropic interlayer interactions. Simple inspection
of Figure 1 leads to 

\begin{equation}
 \label{eq:40}   
   {W}_{int} = 4{J}_{1}({\bf A}\cdot {\bf \Gamma} +
                   {\bf B}\cdot {\bf \Delta}) 
             + 4{J}_{2}({\bf A}\cdot {\bf \Delta} +
                   {\bf B}\cdot {\bf \Gamma}).
\end{equation}
This form is somewhat more involved than the one invoked by Thio et al.
\cite{5} but will eventually lead to the same physical picture.

We must now reformulate the strategy of Section 3 by introducing two
pairs of variables

\begin{eqnarray}
 \label{eq:41}
 &&
 {\bf m}_{1}= \frac{1}{2s}({\bf A}+{\bf B}),\hspace*{1cm}
 {\bf n}_{1}= \frac{1}{2s}({\bf A}-{\bf B}),
            \nonumber
            \\
  &&
  {\bf m}_{2}= \frac{1}{2s}({\bf \Gamma}+{\bf \Delta}),\hspace*{1cm}
  {\bf n}_{2}= \frac{1}{2s}({\bf \Gamma}-{\bf \Delta}),
\end{eqnarray}
which would satisfy two identical copies of the 2D nonlinear $\sigma$
model derived in Section 3 if the interlayer interaction (40) were
neglected. The latter induces a coupling between the two copies which
is especially simple if we invoke the strong inequalities (38).
Then the magnetizations ${\bf m}_{1}$ and ${\bf m}_{2}$ are not directly
affected by the interlayer coupling; i.e.,

\begin{eqnarray}
 \label{eq:42} 
  && 
  {\bf m}_{1} = \frac {\delta}{4}[\,{\bf n}_{1}\times
            (\dot{\bf n}_{1} + {\bf d} - {\bf n}_{1}\times{\bf h})\,],
          \nonumber
          \\
  &&
  {\bf m}_{2} = \frac {\delta}{4}[\;{\bf n}_{2}\times
            (\dot{\bf n}_{2} + {\bf d} - {\bf n}_{2}\times{\bf h})\,],        
\end{eqnarray}
but the staggered moments ${\bf n}_{1}$ and ${\bf n}_{2}$ satisfy
a coupled dynamics described by the total Lagrangian

\begin{eqnarray}
 \label{eq:43}
     &&
     L = L_{0} - V ,         
                           \nonumber
                           \\
     &&
     L_{0} = L_{01} + L_{02} , \hspace*{1cm}  V = V_{1} + V_{2} + V_{12} ,
\end{eqnarray}
where $L_{01}$ and $L_{02}$ are two identical copies of the free
Lagrangian of equation (30) applied for ${\bf n}={\bf n}_{1}$ and
${\bf n}={\bf n}_{2}$, respectively ,$V_{1}$ and $V_{2}$ are similar
copies of the 2D potential, and

\begin{equation}
 \label{eq:44} 
 V_{12} = \rho^{2}({\bf n}_{1} \cdot {\bf n}_{2}), \hspace*{1cm}
          \rho^{2} \equiv \frac{ 8(J_{1} - J_{2}) }{ \delta^{2}J } ,
\end{equation}
is an effective interlayer potential.

The further inequality $J_{1}>J_{2}$ implied by the notation of equation
(44) is simply an assumption consistent with phenomenology.
To illustrate this assumption we consider the ground-state configuration(s)
at zero external field. In this special case the absolute minimum of the
total potential $V$ of equation (43) is achieved when each term
$V_{1}, V_{2}$ or $V_{12}$ assumes its least possible value.
Specifically, ${\bf n}_{1}=-{\bf n}_{2}$ and

\begin{eqnarray}
 \label{eq:45}
 && 
 {\bf n}_{1} = \mp {\bf e}_{b} , \hspace*{1cm}
 {\bf m}_{1} = \pm \frac{\delta}{4}d{\bf e}_{c} ,
                              \nonumber 
                              \\
 &&
 {\bf n}_{2} = \pm {\bf e}_{b} , \hspace*{1cm}        
 {\bf m}_{2} = \mp \frac{\delta}{4}d{\bf e}_{c} ,
\end{eqnarray}
The spin configuration that corresponds to the upper sign is depicted
in Figure 3, whereas the second ground state is obtained by reversing
the sign of all spins and carries the same energy. In either case, 
the average total magnetization 
${\bf m}= \frac{1}{2}({\bf m}_{1}+{\bf m}_{2})$ vanishes and thus
explains the term ``covert weak ferromagnet'' often employed to
describe $La_{2}CuO_{4}$. More involved spin configurations arise
in the presence of external fields and are described in Section 4.1 
and 4.2.

\begin{figure}
\centerline{\hbox{\psfig{figure=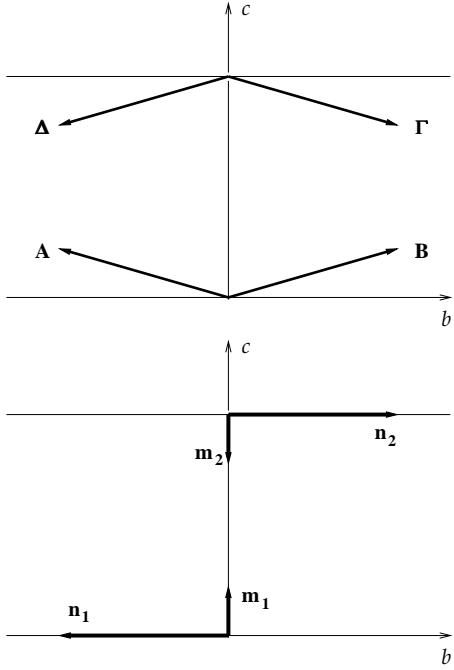,width=6cm}}}
\vskip0.5pc
\caption{Schematic representation of the ground-state configuration
given by equation (45) with the upper sign. A second (degenerate) 
ground state is obtained by
reversing the signs of all spins. The canting angle and the corresponding
magnitude of ${\bf m}_{1}$ and ${\bf m}_{2}$ are greatly exaggerated for
purposes of illustration.}
\end{figure}

This general description of the effective 3D dynamics is completed with
three remarks :

(a) One would think that a more efficient formulation could be obtained
in terms of the Dzyaloshinskii combination of fields \cite{2}

\begin{eqnarray}
 \label{eq:46}
 &&
 {\bf m} = \frac{1}{2}({\bf m}_{1}+{\bf m}_{2}),\hspace*{1cm}
 {\bf n} = \frac{1}{2}({\bf n}_{1}+{\bf n}_{2}),
              \nonumber
              \\
 &&           
 {\bf p} = \frac{1}{2}({\bf m}_{1}-{\bf m}_{2}),\hspace*{1.13cm}
 {\bf q} = \frac{1}{2}({\bf n}_{1}-{\bf n}_{2}),
\end{eqnarray}
which arise naturally in a proper four-sublattice description of 
orthoferrites \cite{26}. The latter are conventional weak ferromagnets
with strong interlayer couplings which lead to 
$|{\bf m}|,|{\bf p}|,|{\bf q}| \ll |{\bf n}|$. One can then show that the
effective low-frequency dynamics may be formulated with a single 
order parameter, 
the total staggered moment ${\bf n}$, and the remaining auxiliary fields
${\bf m},{\bf p}$ and ${\bf q}$ are expressed in terms of ${\bf n}$
through algebraic relations similar to equation (42). However, such
a formulation becomes singular when the interlayer exchange constants
are much weaker than the intralayer one, as is assumed in equation (38).
As a result, the current formulation in terms of two coupled order
parameters ${\bf n}_{1}$ and ${\bf n}_{2}$ is more suitable.

(b) The assumed strong inequalities (38) also imply that gradient terms
of any kind produce negligible corrections to the effective interlayer
coupling. Therefore the homogeneous 3D dynamics described by equations
(42)-(44) may be generalized to a complete continuum field theory
simply by extending the free Lagrangians $L_{01}$ and $L_{02}$ to include
2D spatial gradients according to equation (27) applied for
${\bf n}={\bf n}_{1}$ and ${\bf n}={\bf n}_{2}$, respectively.
The resulting field theory is essentially 2D and the only trace
of 3D dynamics is the bilayer coupling (44).

(c) The interlayer potential (44) vanishes in the tetragonal phase
($J_{1}=J_{2}$). Hence finer anisotropic couplings as well as quantum
fluctuations are necessary to explain spin ordering \cite{30} in, say, 
$Sr_{2}CuO_{2}Cl_{2}$ which remains tetragonal down to temperatures
as low as 10 K \cite{31}.
A field theoretical description of these finer couplings is not available
at present and thus the isotropic 2D nonlinear $\sigma$ model
is the main tool for investigation of tetragonal compounds \cite{32}.

\subsection{Weak-ferromagnetic transition}
\label{subsec:weak_fer_trans} 

We return to the problem posed in the concluding paragraphs of
Section 3 and now address it within its proper 3D context. When
a field of strength $h$ is applied along the $c$ axis the total
bilayer potential is given by

\begin{eqnarray}
 \label{eq:47}
   V &=& {\rho}^{2}({\bf n}_{1}\cdot{\bf n}_{2}) + 
          hd(n_{1b} + n_{2b})             
                       \nonumber
                       \\
     & &  +\frac{1}{2}[a^{2}_{1}(n^{2}_{1a} + n^{2}_{2a}) +
         (a^{2}_{2} + h^{2})(n^{2}_{1c} + n^{2}_{2c})],
\end{eqnarray}      
where ${\bf n}_{1}=(n_{1a},n_{1b},n_{1c})$ and
${\bf n}_{2}=(n_{2a},n_{2b},n_{2c})$  are the two order parameters
expressed in Cartesian components. Simple inspection of the potential,
taking into account that $a_{1}<a_{2}$, suggests that its minima are
such that $n_{1c}=0=n_{2c}$. One may then parametrize the remaining
components as $(n_{1a},n_{1b})=(\cos{\Phi}_{1},\sin{\Phi}_{1})$ and
$(n_{2a},n_{2b})=(\cos{\Phi}_{2},\sin{\Phi}_{2})$ to obtain

\begin{eqnarray}
 \label{eq:48}
     V &=& \rho^{2}\cos({\Phi}_{1} - {\Phi}_{2}) + 
           hd(\sin{\Phi}_{1} + \sin {\Phi}_{2})             
                       \nonumber
                       \\
       & & +\frac{1}{2}a^{2}_{1}(\cos^{2}{\Phi}_{1} + \cos^{2}{\Phi}_{2}).
\end{eqnarray}       
This reduced potential reveals an interesting formal analogy to the case
of an easy-axis antiferromagnet with exchange constant ${\rho}^{2}$,
anisotropy $a^{2}_{1}$, and an effective field of strength $hd$ applied
along the easy $b$ axis; even though the actual field points along the
$c$ axis. 

Therefore the search for the minima of (48) follows the familiar
pattern of the conventional spin-flop transition in an easy-axis
classical antiferromagnet, with due attention to the fact that the
relevant order parameters are now the staggered moments and not the
actual spins. The results of this straightforward minimization
problem can be simply stated by introducing the temporary notational
abbreviations
 
\begin{equation}
 \label{eq:49}
   X = \frac{hd}{\rho^{2}}, \hspace*{1cm}
   Y = \frac{a^{2}_{1}}{\rho^{2}},
\end{equation}
and are summarized in the $T=0$ phase diagram of Figure 4 supplemented
by the typical configurations within each phase illustrated in 
Figure 5.
\begin{figure}

\centerline{\hbox{\psfig{figure=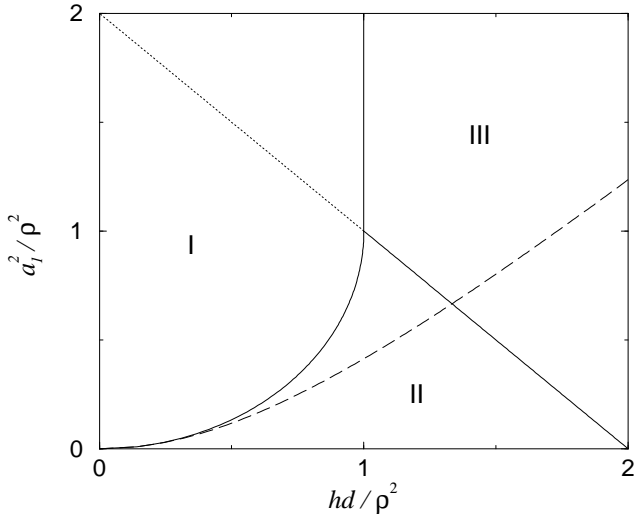,width=8.3cm}}}
\vskip0.5pc
\caption{The $T=0$ phase diagram for a field h parallel to the
$c$ axis. The true critical boundaries are depicted by solid lines,
the limit of local stability of Phase I is shown by a dashed line,
and the same limit of Phase III by a dotted line. The WF transition
in $La_{2}CuO_{4}$ is described by the first-order I : III transition.}
\end{figure}
\begin{figure}

\centerline{\hbox{\psfig{figure=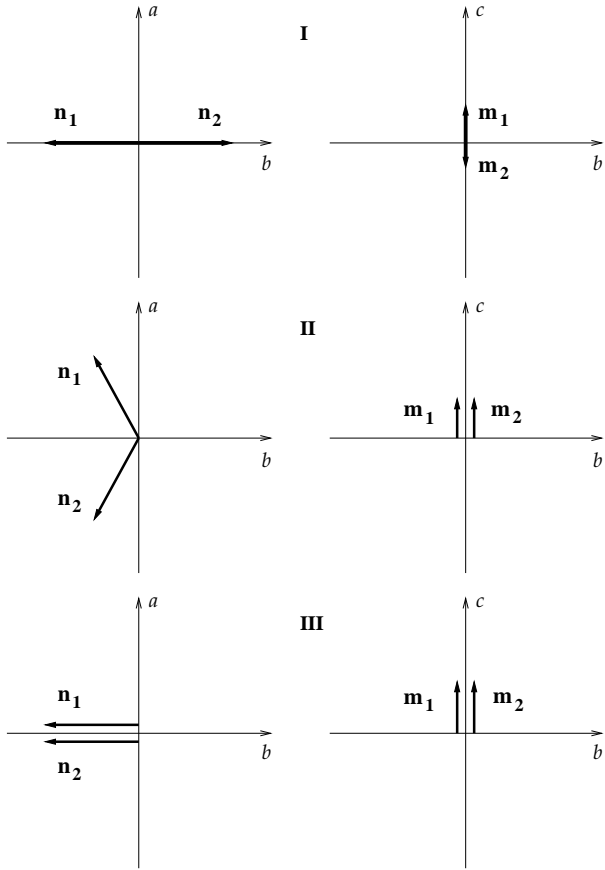,width=8.cm}}}
\vskip0.5pc
\caption{Representative ground-state configurations in Phases I, II, and
III of the $T=0$ phase diagram of Figure 4. In all three cases the net
magnetization ${\bf m} = \frac{1}{2}({\bf m}_{1}+{\bf m}_{2})$ points along
the field direction ($c$ axis).}
\end{figure}
 At zero field $(X=0)$ the ground state is given by equation (45) and
exhibits twofold degeneracy. It is sufficient to consider the configuration
defined by the upper sign in (45) and follow its evolution at
nonvanishing field $h$:

\begin{eqnarray}
 \label{eq:50}
 &      
 {\bf n}_{1} = - {\bf e}_{b},\hspace*{1cm}
 {\bf m}_{1} =\frac{\delta}{4}(h+d){\bf e}_{c},
              \nonumber 
              \\
 &
 {\bf n}_{2} = + {\bf e}_{b}, \hspace*{1cm}           
 {\bf m}_{2} =\frac{\delta}{4}(h-d){\bf e}_{c},   
              \nonumber
              \\ 
        &{\bf m} =\frac{1}{2} ({\bf m}_{1} + {\bf m}_{2}) = 
                 \frac{\delta}{4}h{\bf e}_{c},
\end{eqnarray}
which is depicted in entry I of Figure 5 and exhibits a net moment
${\bf m}$ along the {c} axis whose magnitude increases linearly
with the applied field.

Configuration I remains locally stable until the field crosses the
boundary

\begin{equation}
 \label{eq:51}
   Y = \sqrt {1+X^{2}} -1
\end{equation}
shown by a dashed line in the phase diagram of Figure 4. However
this state becomes metastable at an earlier stage and the true           
critical boundary of Phase I consists of two branches:

\begin{equation}
 \label{eq:52}
 Y = 1 - \sqrt {1 - X^{2}},\hspace*{1cm}  X < 1 ,
\end{equation}
and

\begin{equation}
 \label{eq:53}                                                                 
      X=1 ,\hspace*{1cm} Y > 1 ,
\end{equation}
which are drawn by solid lines in Figure 4 and join a third
critical line
\begin{equation}
 \label{eq:54}                                                                 
    X+Y = 2 ,\hspace*{1cm} Y < 1 ,
\end{equation}
at the ``tricritical'' point $X=1=Y$.

For anisotropies below the tricritical point ($Y<1$) the system would
undergo a first-order transition at the critical line (52) to
enter Phase II characterized by a flopped configuration of the
staggered moments ${\bf n}_{1}$ and ${\bf n}_{2}$ but
magnetizations
${\bf m}_{1}$ and ${\bf m}_{2}$ that are both aligned along the $c$
axis. With further increase of the applied field beyond the critical
boundary (54) a second-order transition occurs and the system enters
Phase III in which both staggered moments are parallel to the 
(negative) $b$ axis.

As we will see in Section 4.3, the parameters of $La_{2}CuO_{4}$ favor
the value $Y=a^{2}_{1} /{\rho}^{2} \approx 2$ and hence the relevant
weak-ferromagnetic (WF) transition is the direct I $\colon$ III transition
that occurs at the critical line (53), or at a critical field
$h=h_{0}$ given by

\begin{equation}
 \label{eq:55}
       h_{0} = {\rho}^{2}/d .
\end{equation}
The WF transition is clearly first order because the boundary of local
stability of Phase I shown by a dashed line in Figure 4 extends
well to the right of the true critical boundary (53).
Similarly the boundary of local stability of Phase III shown by
a dotted line extends well to the left of the true critical boundary.

One may express equation (55) in terms of the original variables
to write

\begin{equation}
 \label{eq:56}
   mH_{0} = s^{2}J_{\perp} ,\hspace*{1cm} J_{\perp} \equiv 4(J_{1}-J_{2}),
\end{equation}
where $H_{0}$ is the critical field, $m$ is the weak moment per
$Cu$ atom at zero field, $s=1/2$ is the spin of a $Cu^{2+}$ ion,
and $J_{\perp}$ is an effective interlayer exchange constant. With
this identification of $J_{\perp}$ equation (56) coincides with
the original estimate of Thio et al. \cite{4}.
In oxygen-doped $ La_{2}CuO_{4+y}$ samples of reduced N\'{e}el
temperature $(T_{N} \sim 240 K)$  the measured critical field is
$H_{0}= 5 T$, $m =2.1 {\times} 10^{-3} {\mu}_{B}$, and thus $J_{\perp}$
is estimated to be $2.5 {\mu}eV$.

The description of the general features of the WF transition is completed
by quoting the explicit ground-state configuration in Phase III:

\begin{eqnarray}
 \label{eq:57}
      &&
      {\bf n}_{1} = -{\bf e}_{b} = {\bf n}_{2} ,
      \nonumber
      \\ 
      &&
      {\bf m} = {\bf m}_{1} = {\bf m}_{2} = 
                \frac {\delta}{4}(d+h){\bf e}_{c} ,
\end{eqnarray}
where the net moment ${\bf m}$ again increases linearly with $h$.
The results for the net magnetization in Phases I and III, given by
equations (50) and (57), are shown by a dashed line in Figure 6
which exhibits  a sudden jump at the critical field $H_{0}=5 T$
due to the first-order nature of the WF transition. One should stress
that this calculation is strictly valid at $T=0$; the magnetization
jump is increasingly smoothed out with rising temperature \cite{4,5}.

\begin{figure}

\centerline{\hbox{\psfig{figure=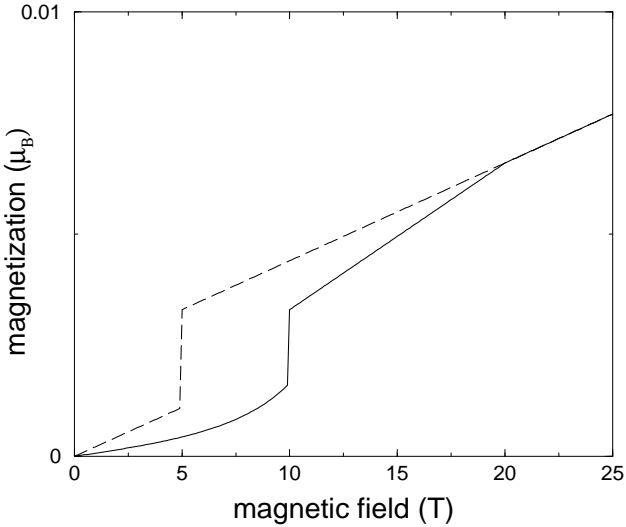,width=8.3cm}}}

\caption{Net magnetization induced by an applied field. The dashed line
corresponds to a field along the $c$ axis and demonstrates the 
WF transition at the critical field $H_{0}=5 T$. The solid line
corresponds to a field along the $b$ axis and displays discontinuities
at the critical fields $H_{1}=10 T$ and $H_{2}= 20 T$ characteristic
of the SF transition. In either case the net magnetization points along
the direction of the applied field.}
\end{figure}

It is now interesting to recalculate the magnon gaps in the presence
of the interlayer coupling and follow them through the WF transition.
If we parametrize ${\bf n}_{1}$ and ${\bf n}_{2}$ by two replicas of
the angular variables (33) the ground-state configuration in 
Phase I is given by 
$({\Theta}_{1}= \frac {\pi}{2}, {\Phi}_{1} = - \frac {\pi}{2})$ and
$({\Theta}_{2}= \frac {\pi}{2}, {\Phi}_{2} = \frac {\pi}{2})$.
Small fluctuations are then studied by introducing the variables
$({\Theta}_{1}= \frac {\pi}{2} - {\theta}_{1},
 {\Phi}_{1} = - \frac {\pi}{2} + {\phi}_{1})$  and
$({\Theta}_{2}= \frac {\pi}{2} - {\theta}_{2},
 {\Phi}_{2} =  \frac {\pi}{2} - {\phi}_{2})$
in the complete Lagrangian (43) and keeping terms that are at most
quadratic :
\begin{eqnarray}
 \label{eq:58}
      L \approx&& \frac{1}{2}
      (\dot {\phi}^{2}_{1} + \dot {\phi}^{2}_{2}
       + \dot {\theta}^{2}_{1} + \dot {\theta}^{2}_{2})
       - {\rho}^{2}({\phi}_{1}{\phi}_{2} + {\theta}_{1}{\theta}_{2})  
      \nonumber
      \\
      & & - \frac{1}{2} (c^{+}_{1}{\phi}^{2}_{1} + c^{-}_{1}{\phi}^{2}_{2}
      + c^{+}_{2}{\theta}^{2}_{1} + c^{-}_{2}{\theta}^{2}_{2}) ,
      \nonumber
      \\
      & &
      \nonumber
      \\
            &&
      c^{\pm}_{1} \equiv a^{2}_{1} \pm hd + {\rho}^{2},
      c^{\pm}_{2} \equiv a^{2}_{2} \pm hd + h^{2} + {\rho}^{2} .
\end{eqnarray}
Standard diagonalization of this quadratic form yields the four
magnon gaps

\begin{eqnarray}
\label{eq:59}
                         &&
  \Omega^{2}_{1^{\pm}} =  a^{2}_{1} + {\rho}^{2}
                            \pm 
                            \sqrt{h^{2}d^{2} + {\rho}^{4}},
                         \nonumber
                         \\
                         &&
                         \nonumber
                         \\
                         &&
  \Omega^{2}_{2^{\pm}} =  a^{2}_{2} + {\rho}^{2}
                            \pm 
                            \sqrt{h^{2}d^{2} + {\rho}^{4}}
                            + h^{2},  
\end{eqnarray}
which reduce to the gaps of equation (36) at zero interlayer coupling
$({\rho}^{2} = 0)$.
The zero-field limit of equation (59) is also interesting and leads
to

\begin{eqnarray}
 \label{eq:60}
&&
    \Omega_{1^{-}} = a_{1},
   \hspace*{2.3cm} \Omega_{2^{-}} = a_{2},
                 \nonumber
     \\ 
     &&
     \nonumber
     \\
     &&               
    \Omega_{1^{+}} = \sqrt {a^{2}_{1} + 2{\rho}^{2}},\hspace*{1cm}
    \Omega_{2^{+}} = \sqrt {a^{2}_{2} + 2{\rho}^{2}}.
\end{eqnarray}
The ``acoustical'' gaps $1^{-}$ and $2^{-}$ do not depend on the
interlayer coupling and correspond to the usual antiferromagnetic (AF)
modes for in-plane and out-of-plane fluctuations, respectively.
The ``optical'' gaps $1^{+}$ and $2^{+}$ are sensitive to the interlayer
coupling and may be said to correspond to exchange (E) modes, in analogy
with a similar distinction made within a proper four-sublattice
formulation of orthoferrites \cite{26}. In the latter case, AF and E 
modes are widely separated thanks to a strong interlayer exchange
interaction that is comparable to the intralayer one. In contrast, a close
proximity of these two types of modes should be expected in $La_{2}CuO_{4}$
because $2{\rho}^{2} \approx a^{2}_{1}$.

The decoupling of AF and E modes suggested by equation (60) no longer
holds in the presence of an applied field, as is apparent in equation (59).
In this respect, it is also useful to follow the gaps beyond the WF
transition. In Phase III the ground-state configuration is given by
${\Theta}_{1}=\frac{\pi}{2}={\Theta}_{2}$, 
${\Phi}_{1}= - \frac{\pi}{2}={\Phi}_{2}$ and small fluctuations lead to
a quadratic Lagrangian similar to (58). The corresponding magnon gaps
are found to be
 
\begin{eqnarray}
 \label{eq:61}    
                     &&   
  \Omega^{2}_{1^{-}} = a^{2}_{1} + hd - 2{\rho}^{2}  ,    
        \nonumber
        \\ 
        &&
        \nonumber
        \\
                     &&
  \Omega^{2}_{2^{-}} = a^{2}_{2} + hd + h^{2} - 2{\rho}^{2}  ,
        \nonumber
        \\
        &&
        \nonumber
        \\ 
                     && 
  \Omega^{2}_{1^{+}} = a^{2}_{1} + hd  ,
        \nonumber
        \\
        &&
        \nonumber
        \\
                     &&
  \Omega^{2}_{2^{+}} = a^{2}_{2} + hd + h^{2}  , 
 \end{eqnarray}
where the role of acoustical and optical modes is clearly interchanged.
To be sure, equation (59) is valid for $h<h_{0}$ and equation (61)
for for $h>h_{0}$. The spectrum exhibits a discontinuity at $h=h_{0}$
thanks again to the first-order nature of the WF transition.

The calculation of dynamic susceptibilities can be effected by the method 
briefly outlined in the concluding paragraph of Section 3 within the simpler
2D context. We do not present here explicit results in order to avoid
further proliferation of algebraic formulas.
But such results can be provided should the need arise.

\subsection{Spin-flop transition}
\label{subsec:sp_fl_tran}

The case of a field applied along some direction in the basal plane
is equally interesting. In particular, when the field is precisely aligned
with the $b$ axis, an unusual spin-flop (SF) transition is reflected in
magnetoresistance measurements \cite{5}.
An important element in the corresponding theoretical analysis is that
the observed in-plane magnon gap is smaller than the out-of-plane gap
$(a_{1}<a_{2})$.

The total bilayer potential in a field ${\bf h}=(0, h, 0)$ is given by

\begin{eqnarray}
 \label{eq:62}
    V &=&  \rho^{2}({\bf n}_{1}\cdot{\bf n}_{2}) - hd(n_{1c} + n_{2c})
         + \frac{1}{2}h^{2}(n^{2}_{1b} + n^{2}_{2b})
        \nonumber
        \\ 
      & & + \frac{1}{2} [ a^{2}_{1}(n^{2}_{1a} + n^{2}_{2a}) +
           a^{2}_{2}(n^{2}_{1c} + n^{2}_{2c})  ]
\end{eqnarray}
Inspite first appearances minimization of the above potential is 
straightforward. The two unit vectors ${\bf n}_{1}$ and
${\bf n}_{2}$ are parametrized in terms of two sets of angular
variables $({\Theta}_{1},{\Phi}_{1})$ and  
$({\Theta}_{2},{\Phi}_{2})$. One can then show that the minima of (62)
are such that

\begin{equation}
 \label{eq:63}
   \Theta_{1} = \Theta _{2} = \Theta
\end{equation}
for any value of the applied field. However the azimuthal angles
${\Phi}_{1}$ and ${\Phi}_{2}$ display different behavior in three
distinct field regions separated by two critical fields:

\begin{equation}
 \label{eq:64}
     h_{1}=a_{1},\hspace*{1.3cm}
     h_{2}=\frac{1}{d} (a^{2}_{2} + 2\rho^{2} - a^{2}_{1}).
\end{equation}
For $h<h_{1}$ the ground-state configuration is illustrated in the
first entry of Figure 7. The staggered moments are both contained
in the $bc$ plane and cant toward the $c$ axis with which they form an angle
${\Theta}$ given by

\begin{figure}

\centerline{\hbox{\psfig{figure=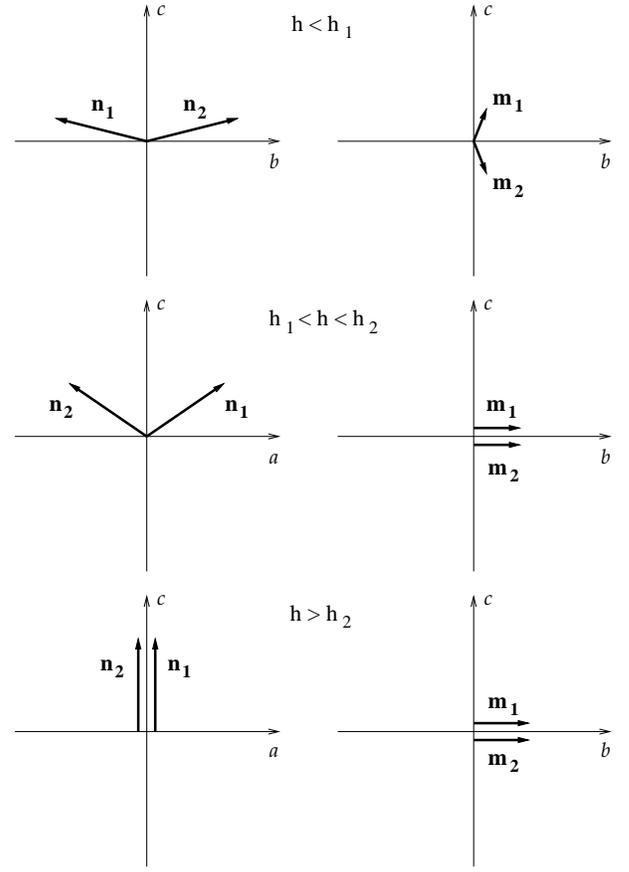,width=8cm}}}
\vskip0.5pc
\caption{Representative ground-state configurations in the three distinct
field regions that characterize the SF transition. In all three cases
the net magnetization ${\bf m}=\frac{1}{2}({\bf m}_{1}+{\bf m}_{2})$ points
in the direction of the applied field ($b$ axis).}
\end{figure}

\begin{equation}
 \label{eq:65}
   \cos {\Theta}=\frac {hd}{a^{2}_{2} + 2{\rho}^{2} - h^{2}}.                    \end{equation}
Accordingly the net magnetization points along the direction of the
applied field, namely
\begin{equation}
 \label{eq:66}
 {\bf m}=\frac {1}{2}({\bf m}_{1} + {\bf m}_{2}) =
         \frac {\delta}{4}(d + h\cos{\Theta})\cos{\Theta}\,{\bf e}_{b}.
\end{equation}
For $h_{1}<h<h_{2}$ the staggered moments flop into the $ac$ plane 
and form an angle with the $c$ axis given by
\begin{equation}
 \label{eq:67}
   \cos{\Theta}=\frac {hd}{a^{2}_{2} + 2{\rho}^{2} - a^{2}_{1}},               
\end{equation}
while the corresponding net magnetization is

\begin{equation}
 \label{eq:68}
  {\bf m} = {\bf m}_{1} = {\bf m}_{2} = 
            \frac{\delta}{4}(h +d\cos{\Theta}){\bf e}_{b}.
\end{equation}
Finally, for $h>h_{2}$, both staggered moments are aligned
with the $c$ axis but the net magnetization
\begin{equation}
 \label{eq:69}
  {\bf m} = {\bf m}_{1} = {\bf m}_{2} = \frac{\delta}{4}(d + h){\bf e}_{b}
\end{equation}
continues to point along the field direction. This picture should be 
completed with the remark that the calculated sharp SF transition
at the critical fields $h_{1}$ and $h_{2}$ is smoothed out when the
direction of the applied field departs from the $b$ axis.

The preceding description of the SF transition confirms the theoretical
analysis of Thio et al. which was in turn shown to be consistent with
experiment \cite{5}.
In particular, the critical fields (64) agree with those of reference
\cite{5} if we adopt the minimal choice $a_{1}=d$ and identify the 
effective interlayer exchange constant $J_{\perp}$ as in equation (56).
The net magnetization calculated from equations (66), (68) and (69) is
depicted by a solid line in Figure 6 and displays characteristic
discontinuities at the observed critical fields $H_{1}=10 T$ and
$H_{2}=20 T$. A minor difference in the overall scale of Figure 6
with the corresponding result of reference \cite{5} is apparently
due to a slightly different choice of parameters, as discussed in the
following subsection.

\subsection{Rationalized units and constants}
 \label{subsec:rat_units_const}
 
Our purpose here is to demonstrate how to efficiently use the rationalized
formulas derived throughout this paper, rather than to analyze in depth the
available experimental data. Such an analysis is complicated by the fact
that actual experiments have been performed on samples with varying
oxygen doping. Pure $La_{2}CuO_{4}$ samples have been available \cite{1,32}
 and exhibit magnetic order below the N\'eel temperature
$T_{N}=325 K$. However the most complete set of magnetic measurements was
 obtained on oxygen-doped $La_{2}CuO_{4+y}$ with reduced N\'eel
temperature \cite{4,5}.
Hence our demonstration will be based on the latter measurements but
could be extended to pure samples in a straightforward manner.

We begin with a parameter-free theoretical prediction based on the fact that
the first critical field in equation (64) and the zero-field ${1^{-}}$ gap
in equation (60) are both equal to $a_{1}$. In physical units this equality
reads $g_{m}{\mu}_{B}H_{1}={\Omega}_{1^{-}}$ where $g_{m}=2.2$ is the
gyromagnetic ratio and ${\mu}_{B}$ the Bohr magneton. 
Hence, if we use the measured
critical field $H_{1}=10.5 \pm 1 T$, the predicted in-plane gap
${\Omega}_{1^{-}}=1.33 \pm 0.12 meV$ is consistent with the measured
$1.1 \pm 0.3 meV$. For simplicity we adopt in the following  the rounded
critical field value $H_{1}=10 T$ which leads to 
${\Omega}_{1^{-}}= 1.27 meV$.

Now the theoretical zero-field weak moment per $Cu$ atom is
$m={\delta}d/4$ or, in physical units, $m=sg_{m}{\mu}_{B}{\delta}d/4$
which should be compared to a measured value 
$2.2\times 10^{-3}{\mu}_{B}$ to yield ${\delta}d=8 \times 10^{-3}$.
This is an estimate of the DM anisotropy recalling that 
${\delta}d=2D/J$. However a more convenient framework is obtained
by exploiting the scale parameter ${\delta}$ to $\em{define}$ 
rationalized units such that $d\equiv1$, and thus
${\delta}=8 \times 10^{-3}$, as anticipated by the discussion of 
Section 3. For the moment we restrict attention to the minimal model
for which $a _{1}= d\equiv 1$. Hence the theoretical critical field 
$h_{1}=a_{1}=1$ sets the rationalized field unit equal to the
measured $H_{1}=10 T$,  and the theoretical in-plane gap
${\Omega}_{1^{-}}=a_{1}= 1$ sets the unit of frequency at $1.27 meV$.
Then the measured critical field  for the WF transition $H_{0}=5 T$
is translated into $h_{0}=1/2$ rationalized units and thus equation
(55) reads $h_{0}={\rho}^{2}/d={\rho}^{2}=1/2$ which provides a rationalized 
estimate of the interlayer coupling. Finally we consider the 
theoretical critical field $h_{2}$ of equation (64) which may now be
applied with $a_{1}=d=1$ and $2{\rho}^{2}=1$ to yield $h_{2}=a^{2}_{2}$
in rationalized units or $H_{2}=10a^{2}_{2} T$ in physical units.
Comparing this prediction to the measured critical field
$H_{2}= 20 T$ we find that $a_{2}=\sqrt{2}$.

To summarize, all theoretical formulas may be applied with
rationalized parameters
\begin{equation}
 \label{eq:70}
 \delta = 8\times 10^{-3};\hspace*{0.3cm} a_{1}=d \equiv 1,
\hspace*{0.3cm} a_{2}= \sqrt{2},\hspace*{0.3cm}   {\rho}^{2}= \frac{1}{2},     \end{equation}
supplemented by the stipulation that the physical unit for
frequency be $1.27 meV$, for field $10 T$, and for magnetization
$sg_{m}{\mu}_{B}=1.1 {\mu}_{B}$.

The $\em{predicted}$ values for the acoustical gaps
${\Omega_{1^{-}}}=1.27a_{1}=1.27 meV$ and
${\Omega_{2^{-}}}=1.27a_{2}=1.8 meV$ are marginally consistent with
the observed $1.1 \pm 0.3 meV$ and $2.5 \pm 0.5 meV$.
In fact, full consistency would be restored if we had
included error bars in our analysis \cite{5} instead of the
conveniently rounded input values for the critical fields and
the magnetization actually used in our demonstration.
Furthermore the optical gaps of equation (60) are now predicted to be
${\Omega_{1^{+}}}=1.27 \sqrt{a^{2}_{1}+ 2{\rho}^{2}}=1.8 meV$
and
${\Omega_{2^{+}}}=1.27 \sqrt{a^{2}_{2}+ 2{\rho}^{2}}=2.2 meV$.
We do not know of an experimental determination of the optical gaps. 
Thus we merely note the predicted close proximity of acoustical and
optical gaps, as anticipated in Section 4.1.

Since the unit of frequency is equal to 
$2s{\delta}J={\delta}J=1.27 meV$, the exchange constant is predicted to
be $J=1.27/{\delta}=158 meV$.
Curiously, this classical value of the exchange constant is the same
with the one obtained in Section 3 in relation to the spinwave velocity
$850 meV\AA$ observed on pure $(T_{N}=325 K)$ samples.
But the spinwave velocity on oxygen-doped samples is typically lower
$(\sim 700 meV\AA)$ and thus the currently predicted classical $J$
is somewhat uncomfortably high. One would think that such a discrepancy
can be averted by resorting to a nonminimal model with
$a_{1}\neq d$. Interestingly, a more comfortable value of $J$ can thus be 
obtained but only at the expense of further deterioration (lowering) of
the prediction for the out-of-plane acoustical gap
${\Omega_{2^{-}}}=1.8 meV$ discussed in the preceding paragraph; and vice
versa. 
Putting it differently, current data do not indicate departure
from the minimal model $a_{1}= d$.

The overall consistency is satisfactory in view of the following limiting
circumstances:

(a) The theoretical model is strictly valid for pure $La_{2}CuO_{4}$
and its application to oxygen-doped samples should be viewed with
caution. Unfortunately, a complete set of magnetic measurements
is not available on pure samples probably because the observed 
magnon gaps are nearly doubled \cite{1} and thus the critical fields
required to induce the SF transition become prohibitively intense.

(b) The theoretical calculation is strictly valid at $T=0$. Although
the measured magnetization and critical fields \cite{4,5} have been
extrapolated to $T=0$, other quantities such as the magnon gaps are
typically determined at relatively high temperatures $(T \sim 100 K)$. 

(c) The assumption of strictly localized spins is always in question, and
the further use of a classical description is marginal for this low spin 
value $(s=1/2)$ and low effective lattice dimension (D = 2).
Since different physical quantities are renormalized differently by
quantum fluctuations, including calculated quantum-renormalization
factors into the classical predictions can be confusing \cite{32}.

\section{Conclusion}
 \label{sec:con}

While the phenomenological picture derived by Thio et al. 
\cite{4,5,6} is confirmed by the present analysis, some new elements 
have emerged that may deserve closer attention:

The structure of the magnon gaps is more involved than normally assumed
because of the underlying four-sublattice magnetic ground state.
The calculated acoustical and optical gaps are not widely separated
and hybridization takes place in the presence of external fields.
Therefore study of the field-dependence of the magnon gaps may lead to
further tests of the derived picture.

The description of the isotropic 2D antiferromagnet in terms of a
relativistic nonlinear $\sigma$ model  has already provided interesting results
\cite{15,16} but the presence of anisotropies and an interlayer coupling
are clearly important for a more detailed understanding of the 
magnetic structure of $La_{2}CuO_{4}$. Suffice it to say that the
existence of a finite N\'eel temperature $(T_{N}=325 K)$ is precisely
due to such perturbations. Hence it is conceivable that the field
theoretical framework discussed here may help to address some of
the remaining questions \cite{32}.

The covert nature of weak ferromagnetism in $La_{2}CuO_{4}$ makes
it difficult to directly observe macroscopic magnetic domains.
Nevertheless, even on structurally pure samples, cooling below
$T_{N}$ should produce numerous magnetic domains and antidomains
separated by domain walls that are invisible because the average
magnetization vanishes at zero external field. When a field is
 applied in a direction perpendicular to the $CuO_{2}$ planes
domain walls evolve into magnetic stripes that exhibit
enhanced magnetization over a region of finite width and could
thus become visible. 
Static magnetic stripes are stable for field
strengths smaller than the critical value $H_{0}$ required 
to induce the WF transition, thanks to the restoring force supplied
by the antiferromagnetic interlayer coupling. When the field exceeds
$H_{0}$ stripes are rendered unstable and begin to expand 
steadily in both directions, thus providing a detailed mechanism
for the observed first-order WF transition.

The preceding qualitative picture may be put on a firm
quantitative basis using the derived continuum field theory,
as we hope to demonstrate in a future publication.

\acknowledgements
This work was supported in part by the Austrian Foundation for the
Promotion of Science (P13846-PHY). JC acknowledges financial support
by the Research Center of Crete during an extended visit, and NP is
grateful to George Robinson for technical assistance.

\appendix
\section{nnn in-plane interactions}
\label{sec:nnn_in-pl_int}
Here we consider the modifications of the 2D Hamiltonian of
Section 2 that result from spin interactions along the diagonals
of the $Cu$ plaquettes. Symmetry precludes the existence
of antisymmetric DM anisotropies on such bonds and thus all
nnn contributions to the Hamiltonian may be cast in the form

\begin{equation}
 \label{eq:71}
 W_{nnn}=\frac{1}{2}\sum_{\ll kl \gg} \sum_{i,j}
         G^{ij}_{kl}(S^{i}_{k}S^{j}_{l} +
         S^{j}_{k}S^{i}_{l}) 
\end{equation}
where $\ll kl \gg$ denotes a nnn bond and the symmetric matrices
$G_{kl}=(G^{ij}_{kl})$ are $\em{not}$ assumed to be traceless.
The possible value of $G_{kl}$ are again restricted by symmetry as shown
in Figure 8.
Thus $A$ spins interact with their $A$ neighbors through the exchange
matrices

\begin{eqnarray}
 \label{eq:72} 
      && 
 G_{1}=\left (\begin{array}{ccc}
                  G_{1aa} & 0       & 0      \\
                  0       & G_{1bb} & G_{1bc} \\
                  0       & G_{1bc} & G_{1cc} 
         \end{array} \right ) ,
                 \nonumber
                 \\
                 &&
                 \nonumber
                 \\
                 &&
                 \nonumber
                 \\
      &&
 G_{2}=\left (\begin{array}{ccc}
                  G_{2aa} & 0       & 0      \\
                  0       & G_{2bb} & G_{2bc} \\
                  0       & G_{2bc} & G_{2cc}
          \end{array} \right ) ,
\end{eqnarray}
along the $a$ and $b$ direction, respectively, whereas $B$ spins
interact with exchange matrices

\begin{eqnarray}
 \label{eq:73} 
       && 
 G'_{1}=\left(
            \begin{array}{ccc}
                G_{1aa} & 0                   & 0 \\
                0       & G_{1bb}             & \hspace*{-0.1cm}-G_{1bc} \\
                0       & \hspace*{-0.1cm}-G_{1bc}& G_{1cc} 
             \end{array} 
         \right) ,
                \nonumber
                \\  
                &&
                \nonumber
                \\
                &&
                \nonumber
                \\
       &&    
 G'_{2}=\left(
           \begin{array}{ccc}
                G_{2aa} & 0                   & 0 \\
                0       & G_{2bb}             & \hspace*{-0.1cm}-G_{2bc} \\
                0       & \hspace*{-0.1cm}-G_{2bc}& G_{2cc} 
           \end{array} 
        \right) ,
\end{eqnarray}
which are related to $G_{1}$ and $G_{2}$. A corollary of these
symmetry relations is that the isotropic components of nnn exchange
couplings, given by the traces of the above matrices, are characterized
by two exchange constants which are generally different along the
$a$ and $b$ directions but the same for $AA$ and $BB$ bonds.

\begin{figure}

\centerline{\hbox{\psfig{figure=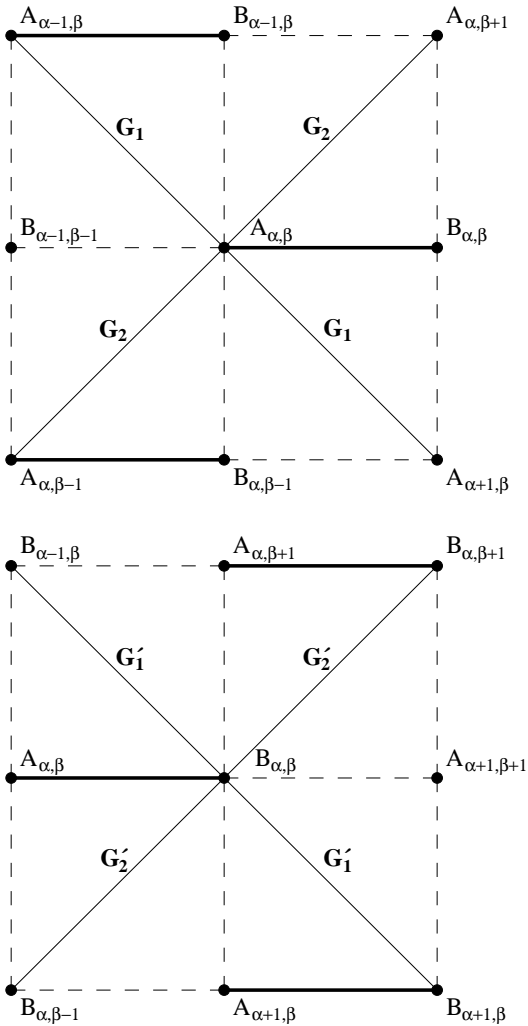,width=7cm}}}
\vskip1pc
\caption{Illustration of the distribution of the matrices $G$ and $G'$
given in the Appendix. These symmetric matrices describe nnn exchange
interactions along the diagonals of the $Cu$ plaquettes which are
parallel to the $a$ and $b$ axes. The remaining conventions
are those of Figure 2.}
\end{figure}

The corresponding modifications of the effective low-frequency
dynamics may be briefly summarized as follows. The traces of the above
matrices introduce an overall additive renormalization of the isotropic
exchange constant $J$.
Similarly the contributions from the diagonal anisotropies submerge
with the anisotropy constants $a_{1}$ and $a_{2}$ already discussed
in the main text. The only new parameter is then introduced by the 
off-diagonal anisotropy, namely

\begin{equation}
 \label{eq:74}
 g = \frac {G_{1bc} + G_{2bc}}{\delta J}, \hspace*{1cm}
 {\bf G}\equiv g(n_{c}{\bf e}_{b} + n_{b}{\bf e}_{c}),
\end{equation}
where we have also defined a vector ${\bf G}$ that depends on the
staggered moment but plays a role similar to the DM vector
${\bf d}=d{\bf e}_{a}$. Then equations (30) generalize to

\begin{eqnarray}
\label{eq:75}
       {\bf m} &=& \frac{\delta}{4}
                  [{\bf n}\times (\dot{\bf n} + {\bf d}
                  + {\bf n}\times {\bf G} - {\bf n}\times {\bf h})] ,
	\nonumber
	\\ 
         & &
        \nonumber
        \\ 
        L_{0}  &=& \frac{1}{2}{\dot{\bf n}}^{2} 
                + \frac{3}{2}g(n^{2}_{b} - n^{2}_{c})\dot n_{a}
                        + {\bf h}\cdot({\bf n}\times{\dot{\bf n}}) ,
	\nonumber
	\\
         & &
        \nonumber
        \\ 
        V      &=& ({\bf h}\times {\bf d})\cdot{\bf n} +
            g\left[
                   h_{c}n_{b} + h_{b}n_{c}
                   - 2({\bf n}\cdot{\bf h})n_{b}n_{c}
             \right] 
         \nonumber
         \\ 
               & &  + \frac{1}{2}({\bf n}\cdot{\bf h})^{2} + 
                \frac{1}{2}(a^{2}_{1}n^{2}_{a} +                               
                            a^{2}_{2}n^{2}_{c})
              + 2g^{2}n^{2}_{b}n^{2}_{c}  ,
\end{eqnarray}
which were actually used throughout our analysis. But the results
presented in the main text were restricted to $g=0$ mainly because
the qualitative picture remains intact and current experimental data
are not sufficiently accurate or detailed to discern a nonvanishing $g$.

\end{document}